\documentclass[%
 reprint,
superscriptaddress,
 amsmath,amssymb,
 aps,
prb,
]{revtex4-2}

\usepackage{graphicx}
\usepackage{multirow}
\usepackage{dcolumn}
\usepackage{bm}
\usepackage{hyperref}
\usepackage{booktabs}
\usepackage{array}
\usepackage[version=4]{mhchem}
\usepackage{makecell} 
\usepackage{makecell}
\usepackage{xcolor}

\begin{document}

\preprint{APS/123-QED}

\title{Machine Learning Study of the Surface Reconstructions of Cu$_{2}$O(111) Surface}

\author{Payal Wadhwa}
 \email{payal.wadhwa@univie.ac.at}
 \email{payal\_wadhwa@iiitvadodara.ac.in}
\altaffiliation{\protect\\Present address:\ Indian Institute of Information Technology Vadodara, Gandhinagar-382028, Gujarat, India}
\affiliation{%
Faculty of Physics and Center for Computational Materials Science, University of Vienna, Kolingasse 14-16, A-1090 Vienna, Austria}

\author{Michael Schmid}
\affiliation{Institute of Applied Physics, TU Wien, A-1040 Vienna, Austria}

\author{Georg Kresse}%
 \affiliation{%
Faculty of Physics and Center for Computational Materials Science, University of Vienna, Kolingasse 14-16, A-1090 Vienna, Austria}
 \affiliation{
VASP Software GmbH, Berggasse 21, A-1090 Vienna, Austria}

\date{\today}

\begin{abstract}
The atomic structure of the most stable reconstructed surface of cuprous oxide (Cu$_{2}$O)(111) surface has been a longstanding topic of debate.\ In this study, we develop on-the-fly machine-learned force fields (MLFFs) to systematically investigate the various reconstructions of the Cu$_{2}$O(111) surface under stoichiometric as well as O- and Cu-deficient or rich conditions, focusing on both ($\sqrt{3}$×$\sqrt{3}$)R30° and (2$\times$2) supercells.\ By utilizing parallel tempering simulations supported by MLFFs, we confirm that the previously described nanopyramidal and Cu-deficient (1×1) structures are the lowest energy structures from moderately to strongly oxidizing conditions.\ In addition, we identify two promising nanopyramidal reconstructions at highly reducing conditions, a stoichiometric and a Cu-rich one.\
Surface energy calculations performed using spin-polarized PBE, PBE+\textit{U}, r$^{2}$SCAN, and HSE06 functionals show that the previously known Cu-deficient configuration and nanopyramidal configurations are at the convex hull (and, thus, equilibrium structures) for all functionals, whereas the stability of the other structures depends on the functional and is therefore uncertain.\ Our findings demonstrate that on-the-fly trained MLFFs provide a simple, efficient, and rapid approach to explore the complex surface reconstructions commonly encountered in experimental studies, and also enhance our understanding of the stability of Cu$_{2}$O(111) surfaces.\

\end{abstract}

\maketitle

\section{Introduction}

Cuprous oxide (Cu$_{2}$O), characterized by the space group Pn$\bar{3}$m, is a \textit{p}-type semiconducting oxide with a wide range of applications in fields such as photovoltaics, solar cells, gas sensors, photocatalysis, dilute magnetic semiconductors, and transparent electronics \cite{meyer2012binary, Ito1997DetailedEO, kakuta2009photocatalytic, chatterjee2016formation, shibasaki2021highly, nilius2024surface, doi:10.1021/acscatal.4c02785, doi:10.1021/acs.jpclett.3c00642,doi:10.1021/acscatal.4c02785, doi:10.1021/acsphotonics.2c01109, doi:10.1021/acsomega.1c04663, https://doi.org/10.1002/pssa.202200887, CHEN2023140089, PhysRevResearch.3.043219, PhysRevMaterials.7.045801}.\ Despite its potential for diverse industrial applications, the precise identification of its most favorable surfaces remained a challenge for an extended period, prompting considerable interest in exploring the surface properties of this material \cite{nilius2024surface}.\ Among the three low-index surfaces, the (111) surface has been identified as the thermodynamically most stable.\ Furthermore, several studies, including those on hydrogen and CO reduction experiments and hydrogen evolution reactions in light-driven experiments to address the catalytic impact of individual (100) and (111) Cu$_{2}$O facets, have highlighted the exceptional catalytic performance of the (111) surface \cite{doi:10.1021/jp904198d, doi:10.1021/jp101617z, chen2022spatiotemporal}.\ This superior performance is attributed to its potentially non-polar and thermodynamically stable nature.\ Additionally, pristine Cu$_{2}$O nanocrystals frequently adopt octahedral shapes, which further underscores the dominance of the (111) crystal facets \cite{doi:10.1021/ja050006u,KUO2010106}.\

The crystal structure of bulk Cu$_{2}$O consists of an fcc Cu sublattice with two tetrahedrally coordinated oxygen atoms per unit cube, such that each Cu atom has a linear coordination to two O atoms at opposite sides (O--Cu--O).\ However, the (111) surface of Cu$_{2}$O is composed of a repeated stacking of electrically neutral O/4Cu/O trilayers \cite{nilius2024surface}.\ Schulz and Cox successfully cleaved single crystals in the Cu$_{2}$O(111) plane and identified two distinct surface terminations: a simple hexagonal (1×1) pattern and a ($\sqrt{3}$×$\sqrt{3}$)R30° unit cell \cite{PhysRevB.43.1610}. Of these two configurations, the (1×1) pattern was attributed to a stoichiometric (ST) surface, prepared by vacuum annealing at 1000 K.\ This surface demonstrated associative O$_{2}$ adsorption only at low temperatures.\ Conversely, the ($\sqrt{3}$×$\sqrt{3}$)R30° reconstruction was observed after multiple adsorption-desorption cycles, inducing oxygen removal from the surface.\ 
However, x-ray photoelectron spectroscopy did not reveal a notable reduction in its oxygen content in comparison to the (1×1) surface \cite{PhysRevB.43.1610}.\ Nevertheless, the ($\sqrt{3}$×$\sqrt{3}$)R30° reconstruction of the Cu$_{2}$O(111) surface initially attributed to oxygen deficiency, prompted extensive experimental and theoretical investigations over the following years \cite{onsten2009atomic,C8CP06023A,PhysRevB.85.035434}.\

Subsequent studies to explore the surface atomic structure of Cu$_{2}$O(111) have predominantly relied on density functional theory (DFT) calculations.\ Previous studies employing the Generalized Gradient Approximation (GGA) have investigated various reconstructions of Cu$_{2}$O(111) surface, including Cu- and O-rich and deficient states \cite{PhysRevB.75.125420,doi:10.1021/acs.jpcc.0c09330}.\ These studies typically commence with a bulk-terminated stoichiometric (111) surface (ST), which is characterized by an interwoven network of Cu-O six-membered rings.\ When cleaving the bulk crystal, this non-polar (stoichiometric) surface exposes an equal number of coordinatively unsaturated Cu and O atoms, which are designated as Cu$_\mathrm{cus}$ and O$_\mathrm{cus}$, respectively.\ The Cu$_\mathrm{cus}$ atoms are positioned at the center of the Cu$_{6}$O$_{6}$ rings and bind to only one subsurface oxygen ion, while O$_\mathrm{cus}$ are located at the corners of the rings.\ The findings indicate that the Cu-deficient (CuD) surface, which lacks the highly undercoordinated Cu$_\mathrm{cus}$ atoms, is thermodynamically the most stable under O-rich conditions.\ 
DFT+U calculations indicate that the CuD surface remains the most stable structure across a wide range of oxygen chemical potentials \cite{doi:10.1021/jp406454c}.\ The relative stability of the ST versus CuD terminations was further examined using hybrid functionals, which revealed that the CuD termination is less favorable due to the increased cost of polarity compensation \cite{nilius2016incorrect,doi:10.1021/acs.jpcc.0c09330,Gloystein_2021}.\  
Scanning Tunneling Microscopy (STM) images of the ($\sqrt{3}$×$\sqrt{3}$)R30° reconstruction reveal triangular protrusions in the superstructure, which prompted the proposal of an alternative surface model \cite{doi:10.1021/acs.jpcc.0c09330, Gloystein_2021, doi:10.1021/acs.jpcc.2c04335}.\ This model entailed the addition of Cu$_{4}$O complexes to the cavities of one-third of the Cu$_{6}$O$_{6}$ rings on a Cu-deficient surface.\ The resulting nanopyramidal reconstruction was found to be energetically favorable when assessed using Heyd-Scuseria-Ernzerhof (HSE) functionals \cite{doi:10.1021/acs.jpcc.0c09330,Gloystein_2021}.\ Nevertheless, finding the structural model of these pyramids remains highly reliant on human expertise.\

The objectives of the present study are summarized below.\ First, we want to determine whether machine-learned force fields (MLFFs) can enumerate low-energy structures for different stoichiometries.\ For simplicity, we decided to train our machine-learned force field on stoichiometric structures using PBE only.\ This is a very pragmatic choice, but easily and conveniently adaptable to other surfaces.\ Once the MLFFs is trained, we combine it with parallel tempering (PT) to identify candidate structures for different stoichiometries.\ We consider the well-studied ($\sqrt{3}$×$\sqrt{3}$)R30° unit cell and the slightly larger (2$\times$2) supercell.\ Subsequently, we incorporated structures from oxygen-deficient and oxygen-rich configurations into the training dataset and generated a second-generation MLFF, and again performed PT simulations for different stoichiometries to evaluate the accuracy of the previously obtained predictions.\ Indeed, the second-generation MLFF confirmed all the structures previously found by the first-generation MLFF, which was trained on stoichiometric structures only.\ The surface energy plots are finally calculated using PBE, PBE+\textit{U}, r$^{2}$SCAN, and HSE06 functionals, as we want to avoid errors associated with MLFFs trained on a limited set of stoichiometries (Supplementary Material Figure S7 \cite{SupplementalMaterial}).\ We conclude by saying that we confirm the structures previously found by theory and experiment, but the main point of the present study is that MLFFs allow one to carry out a largely unbiased search that does not rely on human intuition.\

\section{Computational details}
All calculations are conducted using the Vienna Ab-initio Simulation Package (VASP) \cite{kresse1996efficient,kresse1996efficiency}.\ The Projector-Augmented Wave (PAW) method \cite{blochl1994projector,kresse1999ultrasoft} is used to describe the near-core regions, and electron-electron interactions are treated with Perdew-Burke-Ernzerhof (PBE) \cite{perdew1996generalized}, PBE+\textit{U} \cite{anisimov1991band}, r$^{2}$SCAN \cite{MARQUES20122272,doi:10.1021/acs.jpclett.0c02405}, and HSE06 (25\% exact exchange) \cite{heyd2003hybrid,brothers2008accurate} functionals.\  Standard copper and soft oxygen (O$_{s}$) PAW potentials are employed for all the bulk and surface calculations, while the binding energy of the oxygen molecule for surface energy calculations is determined using hard oxygen PAW potentials (O$_{h}$), see section \ref{sec:bulkphase} for details.\ A Hubbard correction of $U_\mathrm{eff}=4$\ eV is incorporated for d-orbitals of Cu atom for PBE+\textit{U} calculations \cite{C8CP06023A}.\ Bulk Cu$_{2}$O is relaxed with an energy cut-off of 400 eV and a \textit{k}-mesh of (8 $\times$ 8 $\times$ 8), while for surface calculations, a reduced energy cut-off of 300 eV and a \textit{k}-mesh of (2 $\times$ 2 $\times$ 1) is used.\ The optimized lattice constant of bulk Cu$_{2}$O using PBE, PBE+\textit{U}, r$^{2}$SCAN, and HSE06 functionals are 4.306 \AA, 4.289 \AA, 4.252 \AA, and 4.289 \AA, respectively, all in good agreement with the experimental value of 4.27 \AA\ \cite{PhysRevB.25.5929}.\ For surface calculations, a symmetric and sufficiently thick slab of \textit{n} = 11 (O/4Cu/O trilayers) and a vacuum layer of 12 \AA\ is employed to accurately capture the propagation of surface perturbations into the subsurface region and prevent spurious electric fields in the vacuum region.\ 
For calculations using the PBE and PBE+\textit{U} functionals, all surface layers are fully relaxed along the \textit{z}-direction, fixing the in-plane lattice constants to their corresponding bulk value until the force on each atom reaches 0.01 eV/\AA. For r$^{2}$SCAN and HSE06, only electronic optimization was performed.\ The PBE+\textit{U}, r$^{2}$SCAN, and HSE06 structures were obtained by scaling all atomic positions by a uniform scaling constant corresponding to the ratio of the desired and the  PBE lattice constant.\ In the phase diagram, the oxygen chemical potential for bulk transitions are calculated with an energy cut-off of 300 eV with the optimized lattice constants mentioned above.\ Atomic structures are plotted using the VESTA visualization package \cite{https://doi.org/10.1107/S0021889811038970}.

To develop an interatomic potential or force field for the Cu$_{2}$O(111) surface, we employed a kernel-based machine learning model utilizing the on-the-fly force field strategy, which is fully integrated in VASP \cite{jinnouchi2019fly,jinnouchi2020fly}.\ In order to accurately compute thermodynamic properties at elevated temperatures, it is crucial to sample the configuration space as broadly and efficiently as possible.\ The on-the-fly training approach offers an effective solution by significantly reducing the need for first-principles (FP) calculations.\ This method achieves approximately 99\% reduction in FP calculations compared to ``naive" ab initio molecular dynamics (AIMD) or direct DFT-based sampling of equivalent simulation time and configurational diversity, since FP calculations are performed only when the estimated Bayesian uncertainty of the MLFFs becomes notably large. Here, ``Bayesian uncertainty” denotes the square root of the posterior variance in the predicted atomic forces obtained from the Gaussian process regression model \cite{bartok2010gaussian,klawohn2023gaussian}, which serves as a measure of the model’s uncertainty for a given configuration.\ Consequently, computational time during training is reduced by nearly a factor of 100, allowing for extensive configurational space sampling on the picosecond timescale.\ If the training temperatures are sufficiently high (here,  up to 1400 K, 93\% of the experimental melting point temperature), this approach systematically and efficiently captures sufficient configurational space, including high-temperature thermal fluctuations during molecular dynamics (MD) simulations.\ The MLFF was trained on-the-fly with the PBE functional without spin-polarization, during MD simulations in the NVT ensemble, with the temperature controlled by a Langevin thermostat.\ The initial structures for these MD simulations were obtained by relaxing the stoichiometric (111) surface.\ A minimum of 3000 local reference configurations (also called kernel basis functions) were used for both Cu and O atoms.\ The radial and angular cut-offs to represent the local environment of each atom were set to 8 \AA\ and 5 \AA, respectively, with a Gaussian width of 0.5 \AA\ to broaden the atomic distributions.\ To further describe the atomic environment, 12 spherical Bessel functions were used for radial descriptors, and spherical harmonics up to \textit{l} = 3 were employed.\ The final force fields were obtained by refitting the energies and forces using a kernel-based method.\

\section{Results and Discussions}
\begin{figure*}[t]
\centering
\includegraphics[width=\textwidth]{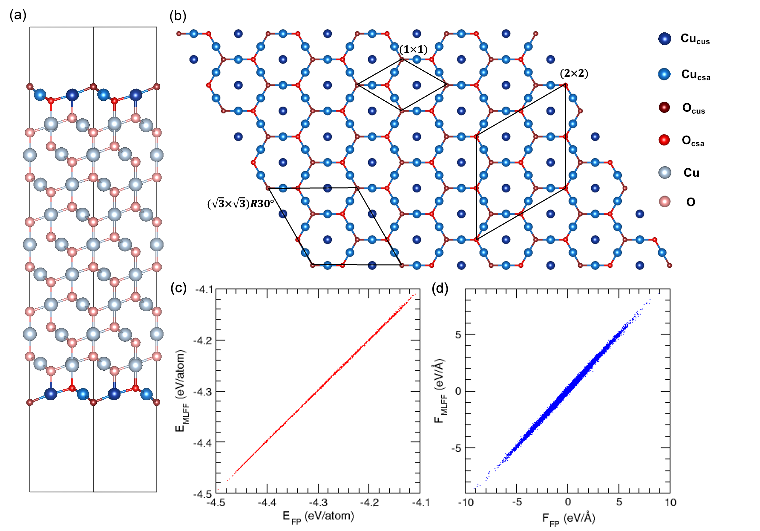}
\caption{\label{figure1}Crystal structure of the bulk-terminated stoichiometric Cu$_{2}$O(111) surface:\ (a) Side view and (b) Top view, with atom color codes provided in the top right corner.\ Darker spheres represent atoms on the surface layer, while lighter spheres indicate atoms beneath the surface layer.\ For better visualization, the sphere sizes of Cu and O atoms below the surface layer have been enlarged.\ Navy blue and ocean blue spheres denote coordinatively unsaturated and saturated Cu atoms, respectively, while maroon and red spheres represent the coordinatively unsaturated and saturated O atoms, respectively.\ Panel (c) and (d) depict the comparison of energies $E$ and forces $F$ from the training dataset, computed using machine-learned force fields (MLFF) and first-principles (FP) calculations, respectively.}
\end{figure*}

To develop the machine-learned force fields (MLFFs) for the Cu$_{2}$O(111) surface, we utilized a symmetric, 11-layers, bulk-terminated stoichiometric ($\sqrt{3}$×$\sqrt{3}$)R30° surface model composed of 132 Cu and 66 O atoms, as depicted in Figure \ref{figure1}(a).\ The system was gradually heated from 50 K to 1400 K over 200 ps, followed by a constant-temperature MD simulation at 1400 K for an additional 200 ps.\ Subsequently, the structure was cooled from 1400 K to 300 K over 200 ps and then subjected to another constant-temperature MD simulation at 1400 K for 200 ps.\ This training process generated a total of 1645 configurations.\ The energies and forces computed using the MLFFs and first-principles (FP) for the training set are plotted in Figures \ref{figure1}(c) and (d), respectively.\ The data points align closely with the parity line (\textit{y} = \textit{x}), indicating the accuracy of the generated MLFFs.\ In addition, we constructed a test-set of 200 structures by sampling snapshots from constant temperature MD runs at 1200 K and 1400 K.\ The root mean square errors (RMSE) for the energies and forces for the training and test-data sets are listed in Table \ref{tab:rmse} and are nearly identical, further confirming the reliability of the MLFFs.\ The structural diversity of the training datasets are also evaluated using principal component analysis, further supporting the robustness of the MLFFs (Supplementary Material S1 \cite{SupplementalMaterial}).\ The resulting force fields were then employed to explore various surface reconstructions of the Cu$_{2}$O(111) surface.\

\begin{table}[h]
\centering
\caption{Root mean square errors (RMSE) for energies (meV/atom) and forces (meV/Å) of the training and test sets.}
\label{tab:rmse}
\begin{tabular}{lcc}
\hline
\textbf{Dataset} & \textbf{\makecell{Energy RMSE \\ (meV/atom)}} & \textbf{\makecell{Force RMSE \\ (meV/\AA)}} \\
\hline
Training set & 1 & 52 \\
Test set     & 1 & 64 \\
\hline
\end{tabular}
\end{table}

\begin{figure*}
\includegraphics[width=\textwidth]{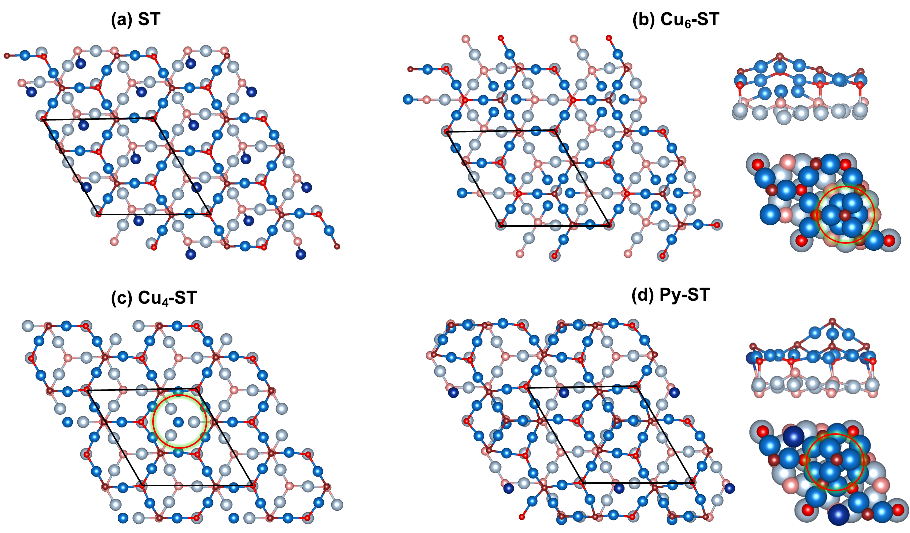}
\caption{\label{figure2}Atomic structures of the ($\sqrt{3}$×$\sqrt{3}$)R30° Cu$_{2}$O(111) surface showing (a) the bulk-truncated stoichiometric termination after relaxation (ST), (b) a 6-atom Cu cluster (Cu$_\mathrm{6}$-ST), (c) a 4-atom Cu cluster (Cu$_\mathrm{4}$-ST), and (d) a stoichiometric nanopyramidal configuration (Py-ST).\ All representations include the subsurface layer, with light blue and red spheres representing Cu and O atoms, respectively.\ For (b) and (d), the left panels display the top view, while the upper and lower right panels show the side view and the space-filling model of the respective configurations, respectively.\ In (b), (c), and (d) red-green circles highlight the formation of the Cu cluster and nanopyramid, respectively.\ Atom color code follows that of Figure \ref{figure1}.}
\end{figure*}

 It is important to note that in the stoichiometric configuration obtained from the bulk-terminated (111) surface, the Cu$_\mathrm{cus}$ ions are located at the center of the ring [Figure \ref{figure1}(b)].\ However, previous theoretical studies have indicated that these ions would deviate slightly from the center, breaking the symmetry \cite{nilius2024surface,doi:10.1021/acs.jpcc.0c09330}.\ To investigate this, we began with the original symmetric configuration and conducted an MD simulation assisted by MLFFs at 300 K for 500 steps.\ It is found that the Cu$_\mathrm{cus}$ ions get displaced from the center rapidly, thus breaking the symmetry, as shown in Figure \ref{figure2}(a).\ This is consistent with previous studies \cite{nilius2024surface,doi:10.1021/acs.jpcc.0c09330}.\ This configuration was further optimized using MLFFs and then using FP calculations.\ This arrangement is designated as ST configuration as shown in Figure \ref{figure2}(a).\

In order to investigate potential surface reconstructions of the Cu$_{2}$O(111) surface for both the ($\sqrt{3}$×$\sqrt{3}$)R30° unit cell and (2$\times$2) supercells under stoichiometric as well as O- and Cu-deficient or enriched conditions, we employed a global energy optimization algorithm known as Parallel Tempering (PT) \cite{B509983H} assisted by MLFFs, by varying the temperatures from 550 K to 1120 K.\ The details of the resulting favorable reconstructions for ($\sqrt{3}$×$\sqrt{3}$)R30° unit cell and (2$\times$2) supercells are provided in sections \ref{A} and \ref{B}.\ Given that prior theoretical studies have indicated a preference for Cu-deficient or O-rich configurations \cite{nilius2024surface}, we put greater emphasis on exploring various combinations of these stoichiometries, as summarized in Table \ref{Table1}.\

\begin{table*}
\centering
\caption{\centering Various stoichiometries obtained by altering the number of Cu and O atoms relative to the stoichiometric configuration (underlined for both ($\sqrt{3}$×$\sqrt{3}$)R30° and (2×2) supercell).} 
\label{Table1}

\begin{tabular}{l c c c c l}
\toprule
\hspace{0.01cm}\textbf{Configuration} \hspace{0.01cm} & \hspace{0.04cm} \textbf{Size of cell} \hspace{0.04cm}& \hspace{0.04cm} \makecell{\textbf{Number of} \\ \textbf{Cu atoms}} \hspace{0.05cm} & \hspace{0.04cm} \makecell{\textbf{Number of} \\ \textbf{O atoms}} \hspace{0.01cm} & \hspace{0.04cm} \makecell{\textbf{Cu excess} \\ \textbf{per (1×1) cell}} \hspace{0.05cm} & \hspace{0.01cm} \makecell{\textbf{Name of} \\ \textbf{reconstruction}} \hspace{0.01cm} \\

\midrule
\multirow{4}{*}{\centering \textbf{Stoichiometric}} 
    & ($\sqrt{3}$×$\sqrt{3}$)R30° & \underline{132} & \underline{66} & 0 & ST [Figure \ref{figure2}(a)] \& Cu$_\mathrm{6}$-ST [Figure \ref{figure2}(b)] \\  \cmidrule(l){2-6} 
    & ($\sqrt{3}$×$\sqrt{3}$)R30° & 128 & 64 & 0 & Cu$_\mathrm{4}$-ST [Figure \ref{figure2}(c)] \\  \cmidrule(l){2-6} 
    & ($\sqrt{3}$×$\sqrt{3}$)R30° & 136 & 68 & 0 & Py-ST [Figure \ref{figure2}(d)] \\  \cmidrule(l){2-6} 
    & (2×2) & \underline{176} & \underline{88} & 0 & Py-ST$_\mathrm{(2\times2)}$ [Figure \ref{figure4}(a)] \\  
\midrule
\multirow{7}{*}{\centering \textbf{Cu-deficient}}  
    & ($\sqrt{3}$×$\sqrt{3}$)R30° & 130 & 66 & -$\frac{1}{3}$ & $\frac{1}{3}$CuD (Supplementary Figure S3 \cite{SupplementalMaterial}) \\  \cmidrule(l){2-6} 
    & ($\sqrt{3}$×$\sqrt{3}$)R30° & 128 & 66 & -$\frac{2}{3}$ & $\frac{2}{3}$CuD (Supplementary Figure S3 \cite{SupplementalMaterial}) \\  \cmidrule(l){2-6} 
    & ($\sqrt{3}$×$\sqrt{3}$)R30° & 126 & 66 & -1 & CuD [Figure \ref{figure3}(a)] \\  \cmidrule(l){2-6} 
    & ($\sqrt{3}$×$\sqrt{3}$)R30° & 126 & 64 & -$\frac{1}{3}$ & CuD-O$_\mathrm{v_{ss}}$ [Figure \ref{figure3}(c)] \\  \cmidrule(l){2-6} 
    & ($\sqrt{3}$×$\sqrt{3}$)R30° & 134 & 68 & -$\frac{1}{3}$ & Py-CuD [Figure \ref{figure3}(d)] \\  \cmidrule(l){2-6} 
    & (2×2) & 176 & 90 & -$\frac{1}{2}$ & Py-CuD$_\mathrm{(2\times2)}$ [Figure \ref{figure4}(b)] \\  \cmidrule(l){2-6} 
    & (2×2) & 174 & 88 & -$\frac{1}{4}$ & Cu$_\mathrm{6}$-CuD$_\mathrm{(2\times2)}$ [Figure \ref{figure4}(c)] \\  
\midrule
\multirow{3}{*}{\centering \textbf{O-deficient}}  
    & ($\sqrt{3}$×$\sqrt{3}$)R30° & 132 & 64 & +$\frac{2}{3}$ & ST-O$_\mathrm{v_{ss}}$ [Figure \ref{figure3}(b)] \\  \cmidrule(l){2-6} 
    & ($\sqrt{3}$×$\sqrt{3}$)R30° & 130 & 64 & +$\frac{1}{3}$ & $\frac{1}{3}$CuD-O$_\mathrm{v_{ss}}$ (Supplementary Figure S3 \cite{SupplementalMaterial}) \\  \cmidrule(l){2-6} 
    & ($\sqrt{3}$×$\sqrt{3}$)R30° & 134 & 66 & +$\frac{1}{3}$ & Py-O$_\mathrm{v_{ss}}$ [Figure \ref{figure3}(e)] \\  
\bottomrule
\end{tabular}
\end{table*}

\subsection{\label{A} \texorpdfstring{($\sqrt{3}$×$\sqrt{3}$)}R30° surface reconstructions}

First, we considered three stoichiometric configurations:\ the ST configuration having 132 Cu and 66 O atoms, with each of the Cu$_\mathrm{6}$O$_\mathrm{6}$ rings (``cavities'') filled by one undercoordinated Cu atom, as shown in Figure 2(a), along with two additional configurations obtained by removing/adding two Cu atoms and one O atom symmetrically from/to both sides of the ST surface.\ These modifications resulted in compositions of (128 Cu, 64 O) and (136 Cu, 68 O) atoms, respectively.\ Parallel tempering (PT) simulations for the three configurations led to the following lowest energy structures:

\begin{enumerate} 
\item[1.] \textbf{Cu$_\mathrm{6}$-ST} \textit{[Figure\ \ref{figure2}(b)]}:\ For the ST configuration comprising 132 Cu and 66 O atoms, PT simulations yield a structure where the originally undercoordinated three Cu$_\mathrm{cus}$ atoms move toward a common corner point of the three Cu$_{6}$O$_{6}$ cavities, to form a Cu cluster of six atoms.\ The oxygen atom at this corner point moves up and is now situated at the top of the Cu cluster.\ In the resulting structure, the three uppermost Cu atoms have the same coordination to oxygen as in the nanopyramids discussed below.\ The bonding distances between the three upper and the three lower Cu atoms are 2.5 Å, slightly less than that in metallic Cu (2.55 Å), which motivates the designation as a Cu$_\mathrm{6}$ cluster.

\item[2.] \textbf{Cu$_\mathrm{4}$-ST} \textit{[Figure\ \ref{figure2}(c)]}:\ In this stoichiometric configuration comprising 128 Cu and 64 O atoms, achieved by removing two Cu and one O atoms from each side of the ST surface,
two of the three undercoordinated Cu$_\mathrm{cus}$ atoms are eliminated.\ The remaining Cu$_\mathrm{cus}$ atom is retained within one of the three Cu$_{6}$O$_{6}$ cavities; however, the O atom beneath it (the only O neighbor of the Cu$_\mathrm{cus}$ in the truncated bulk structure) disappears.\ Thus, this undercoordinated Cu atom sinks down and forms three strong bonds (2.43 Å with the PBE functional) to the second-layer Cu atoms that have lost an oxygen neighbor.\ Also, the Cu--Cu distances between these three subsurface atoms (2.53 Å) are shorter than in metallic Cu, thus we have a cluster of 4 Cu atoms.

\item[3.] \textbf{Py-ST} \textit{[Figure\ \ref{figure2}(d)]}:\ The last stoichiometric configuration, having 136 Cu and 68 O atoms, is similar to the nanopyramidal structure found previously \cite{nilius2024surface,doi:10.1021/acs.jpcc.0c09330}.\ The key distinction lies in the presence of a Cu$_\mathrm{cus}$ atom, analogous to the ST structure [Figure\ \ref{figure2}(a)], which occupies one of the cavities formed by the Cu$_6$O$_6$ surface rings.\ The nanopyramid itself is also located in one of the cavities.\ The pyramid is best described as a larger Cu$_{2}$O agglomeration, comprising a capping O atom and three Cu atoms in the top layer, with linear O--Cu--O coordination, and also binding to a single Cu atom below the center of the pyramid.\ This Cu atom is close to the position of the undercoordinated Cu$_\mathrm{cus}$ of the bulk-truncated surface, but now binds to one O atom (below) and the three aforementioned Cu atoms.\ 

We also investigated additional stoichiometric configurations containing (124 Cu, 62 O) and (140 Cu, 70 O) atoms, obtained by further removal or addition of Cu and O atoms to the ST configuration.\ Since there is no experimental evidence for these structures and their energies are unfavorable, they are presented in Supplementary Material S2 \cite{SupplementalMaterial}.\
\end{enumerate}

\begin{figure*}
\includegraphics[width = \textwidth]{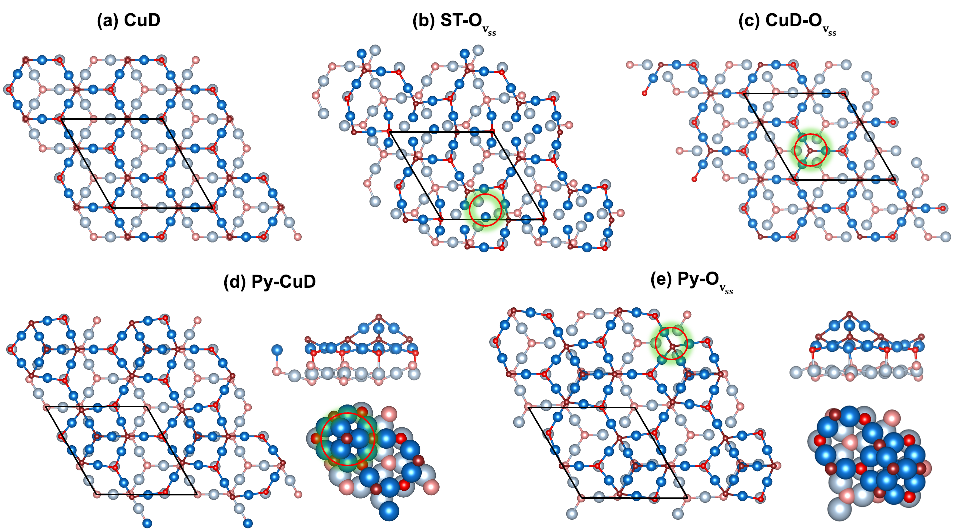}
\caption{\label{figure3}Atomic structures of the ($\sqrt{3}$×$\sqrt{3}$)R30° Cu$_{2}$O(111) surface showing (a) Cu-deficient surface (CuD), (b) oxygen vacancy stoichiometric surface (ST-O$_\mathrm{v_{ss}}$), (c) oxygen vacancy Cu-deficient surface (CuD-O$_\mathrm{v_{ss}}$), (d) Cu-deficient nanopyramidal reconstruction (Py-CuD), and (e) oxygen vacancy nanopyramidal configuration (Py-O$_\mathrm{v_{ss}}$).\ 
For (d) and (e), the left panels display the top view, while the upper and lower right panels show the side view and the space-filling model of the nanopyramidal configuration, respectively.\ In (b), (c), and (e), the subsurface oxygen vacancy is highlighted by a red-green circle, while in (d), the same marker indicates the formation of the nanopyramid. The atom color coding corresponds to that of Figure \ref{figure1}.}
\end{figure*}

Subsequently, we simulated various O- and Cu-deficient or enriched states for ($\sqrt{3}$×$\sqrt{3}$)R30° Cu$_{2}$O(111) surface, as described in the following:

\begin{enumerate} 

\item[1.] \textbf{CuD} \textit{[Figure\ \ref{figure3}(a)]}:\ First, we consider a Cu-deficient system, by sequentially removing one, two, and three copper atoms from each side of the symmetric ST configuration, reducing the composition to (130 Cu, 66 O), (128 Cu, 66 O), and (126 Cu, 66 O) atoms.\ In all three cases, PT simulations yield configurations where Cu$_\mathrm{cus}$ atoms are systematically absent from the cavities between Cu$_6$O$_6$ surface rings of one-third, two-thirds, and the entire surface layer, respectively.\ These configurations are designated as $\frac{1}{3}$CuD, $\frac{2}{3}$CuD, and CuD configurations, respectively.\ The resulting CuD configuration is similar to the one reported in previous studies \cite{ doi:10.1021/acs.jpcc.0c09330,Gloystein_2021, dongfang2023understanding}, and is depicted in Figure \ref{figure3}(a), while the $\frac{1}{3}$CuD, $\frac{2}{3}$CuD configurations are presented in the Supplementary Material Figure S3 \cite{SupplementalMaterial}.
\item[2.] \textbf{ST-O$_\mathrm{v_{ss}}$} \textit{[Figure\ \ref{figure3}(b)]}:\ Here, we consider an oxygen-deficient system, by removing one oxygen atom from each side of the symmetric ST configuration (132 Cu and 64 O atoms).\ PT simulations identified the ground state structure as having one O atom absent from one-third of the Cu$_{6}$O$_{6}$ subsurface rings.\ This leads to substantial distortions, breaks the threefold symmetry, and also the Cu$_{6}$O$_{6}$ rings break up.\ The oxygen vacancy [marked by a circle in Figure \ref{figure3}(b)] migrates to the subsurface layer as also observed in previous studies \cite{doi:10.1021/acs.jpcc.0c09330,dongfang2023understanding}.\ This configuration is designated as ST-O$_\mathrm{v_{ss}}$, where the subscript ``$\mathrm{v_{ss}}$" denotes the oxygen vacancy in the subsurface layer.\\
\item[3.] \textbf{CuD-O$_\mathrm{v_{ss}}$} \textit{[Figure\ \ref{figure3}(c)]}: Here, we considered two configurations: (130 Cu, 64 O) and (126 Cu, 64 O), created by removing one Cu and one O atom and three Cu and one O atoms from each side of the ST configuration.\ PT simulations for these configurations identified their ground state structures as follows: In the first configuration, Cu$_\mathrm{cus}$ atoms are absent from the cavities of one-third of the Cu$_{6}$O$_{6}$ surface rings, while in the second configuration, Cu$_\mathrm{cus}$ atoms are absent from the cavities of the entire surface rings.\ In both cases, one O atom is absent from the subsurface layer.\ These configurations are referred to as $\frac{1}{3}$CuD-O$_\mathrm{v_{ss}}$ and CuD-O$_\mathrm{v_{ss}}$, respectively.\ A graphical representation of the CuD-O$_\mathrm{v_{ss}}$ configuration is provided in Figure \ref{figure3}(c), while the $\frac{1}{3}$CuD-O$_\mathrm{v_{ss}}$ configuration is included in the Supplementary Material Figure S3 \cite{SupplementalMaterial}.\ The highlighted red-green circle in the figure illustrates the position of the oxygen vacancy, which is again found in the subsurface layer.\
\item[4.] \textbf{Py-CuD} \textit{[Figure\ \ref{figure3}(d)]}: Here, we investigated another Cu-deficient configuration by adding one Cu and one O atom to each side of the symmetric ST surface, yielding a system with 134 Cu and 68 O atoms.\ PT simulations revealed the formation of a nanopyramidal structure in the cavities of one-third of the Cu$_{6}$O$_{6}$ surface rings.\ The nanopyramid structure is exactly the same as observed in Py-ST [Figure \ref{figure2}(d)].\ It also comprises an apical oxygen on top of three Cu atoms in the top layer and seven Cu atoms in the surface layer.\ The apical oxygen is 1.85 \AA above the three Cu atoms in the top layer (in-plane Cu–Cu distance 2.72 \AA).\ This structure is referred to as Py-CuD.\ When considering the smearing by the finite lateral resolution of the STM, a topographic height of the Cu surface triangle (1.8 \AA) is consistent with the experimentally observed STM height of 1.5 \AA\ for the Cu-deficient nanopyramidal configuration \cite{doi:10.1021/acs.jpcc.2c04335,C8CP06023A}.\ It is to be noted that prior theoretical studies constructed nanopyramidal surface models by incorporating Cu$_{4}$O complexes into a Cu-deficient (CuD) surface \cite{doi:10.1021/acs.jpcc.0c09330}.\ In contrast, our PT simulations autonomously converged to the correct ground-state structure without requiring such preconstructed models.\ 
\item[5.] \textbf{Py-O$_\mathrm{v_{ss}}$} \textit{[Figure\ \ref{figure3}(e)]}: Lastly, we simulated a Cu-rich/O-deficient configuration by introducing one additional Cu atom to each side of the symmetric ST surface, resulting in a system comprising 134 Cu and 66 O atoms.\ PT simulations revealed the emergence of the same nanopyramidal structure as in Py-ST and Py-CuD, with an oxygen vacancy that again migrates to the subsurface layer, as indicated by the highlighted circle in Figure \ref{figure3}(e).\ 

It is noteworthy that further removal or addition of Cu and O atoms did not result in any structures qualitatively different from those discussed above.\ For reference, stoichiometries of (138 Cu, 68 O) and (138 Cu, 70 O) are presented in the Supplementary Material Figure S4 \cite{SupplementalMaterial}.\
\end{enumerate}

\subsection{\label{B} (2×2) supercell reconstructions}
\begin{figure*}
\includegraphics[width = 15cm]{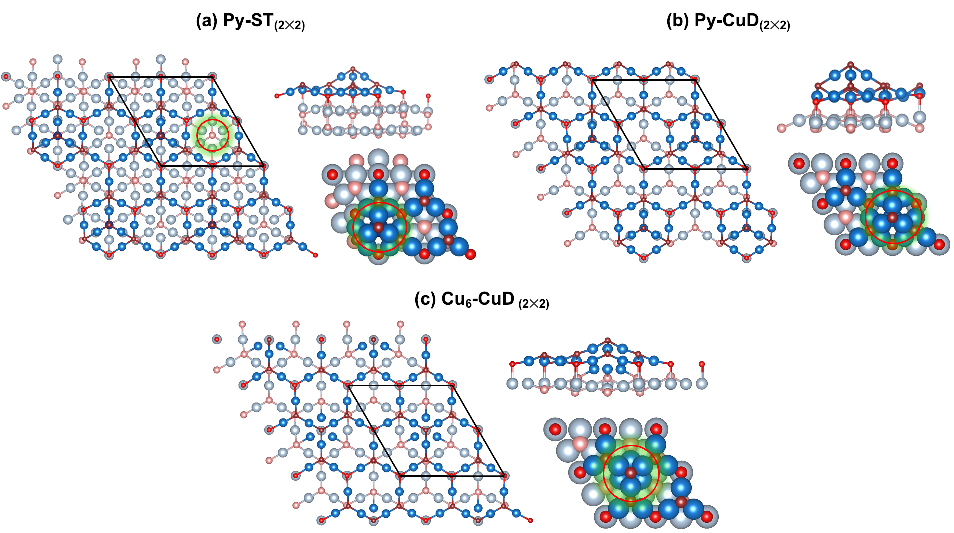}
\caption{\label{figure4}Atomic structures for the (2×2) Cu$_{2}$O(111) surface, showing (a) a stoichiometric nanopyramidal configuration (Py-ST$_\mathrm{(2\times2)}$), (b) a Cu-deficient nanopyramidal configuration (Py-CuD$_\mathrm{(2\times2)}$), and (c) a 6-atom Cu cluster arrangement (Cu$_\mathrm{6}$-CuD$_\mathrm{(2\times2)}$).\ For (a), the layer beneath the subsurface is also shown.\ The left panels display the top view, while the upper and lower right panels show the side view and the space-filling model of the respective configurations, respectively.\ In the left panel of (a), the red-green circle highlights the formation of an oxygen vacancy.\ In the right panels of (a), (b), and (c), the same marker denotes the formation of a nanopyramid and a Cu cluster.\ The atom color coding corresponds to that of Figure \ref{figure1}.}
\end{figure*}

\begin{enumerate} 
\item[1.] \textbf{Py-ST$_\mathrm{(2\times2)}$} \textit{[Figure\ \ref{figure4}(a)]}: To explore potential surface reconstructions for the (2×2) supercell, we started with the stoichiometric configuration (ST) consisting of 176 Cu and 88 O atoms.\ PT simulations revealed a similar nanopyramid as identified in Py-ST and Py-CuD for the ($\sqrt{3}$×$\sqrt{3}$)R30° unit cell.\ While the upper layers of this nanopyramid are exactly as described above for the Py-CuD structure, there is an additional oxygen vacancy rather deep beneath the pyramid, between the first and the second subsurface Cu layer, as shown by the highlighted circle in the top view of Figure \ref{figure4}(a).\

\item[2.] \textbf{Py-CuD$_\mathrm{(2\times2)}$} \textit{[Figure\ \ref{figure4}(b)]}: Here, we simulated a Cu-deficient configuration by adding one O atom to each side of the symmetric (2×2) ST configuration, resulting in a composition having 176 Cu and 90 O atoms.\ PT simulations again revealed a nanopyramidal structure similar to Py-ST$_\mathrm{(2\times2)}$, except the added oxygen atom now fills the vacancy beneath the subsurface layer.\ This structure is referred to as Py-CuD$_\mathrm{(2\times2)}$.

\item[3.] \textbf{Cu$_\mathrm{6}$-CuD$_\mathrm{(2\times2)}$} \textit{[Figure\ \ref{figure4}(c)]}: Subsequently, we simulated another Cu-deficient configuration by removing one Cu atom from each side of the symmetric (2×2) ST configuration, resulting in a composition having 174 Cu and 88 O atoms.\ PT simulations revealed a ground state structure similar to the Cu$_\mathrm{6}$-ST obtained for ($\sqrt{3}$×$\sqrt{3}$)R30° unit cell, where three originally undercoordinated Cu$_\mathrm{cus}$ atoms move toward three fully coordinated Cu$_\mathrm{csa}$ atoms, to form a 6-atom Cu cluster, with an oxygen atom at the tip.\ Therefore, the structure is designated as Cu$_\mathrm{6}$-CuD$_\mathrm{(2\times2)}$.\ Notably, further removal of Cu and O atoms from the (2×2) supercell did not produce any new configurations beyond those identified in the ($\sqrt{3}$×$\sqrt{3}$)R30° unit cell.\  

\end{enumerate}

\subsection{\label{C} Validity of MLFFs for non-stoichiometric configurations}

To ensure the validity of the generated MLFFs for non-stoichiometric configurations, we modified the stoichiometric machine-learned potential by incorporating a set of structures representative of Py-CuD [Figure \ref{figure3}(d)] and Py-O$_\mathrm{v_{ss}}$ [Figure \ref{figure3}(e)], as they inherently possess Cu-deficient and oxygen-deficient (or Cu-rich) structures, respectively.\ To do this, we selected the trajectories of the images of the parallel tempering runs that yielded the lowest energies for 
Py-CuD  and Py-O$_\mathrm{v_{ss}}$.\
From these runs, every 1000th structure from the MD trajectory was chosen \cite{staacke2021role,jung2023machine}.\ 
This gives a total of 500 additional structures for each stoichiometry.\ All these structures were fed in a single calculation to VASP, and a DFT calculation was performed for all structures with a Bayesian uncertainty above a predefined threshold.\ These structures, their DFT energies, and forces were added to the MLFF training dataset (Supplementary Material S4 \cite{SupplementalMaterial}).\ In total 102 additional structures were included, comprising 56 structures for Py-CuD and 46 structures for Py-O$_\mathrm{v_{ss}}$.\ While the original (first-generation) MLFF was unreliable in determining energy differences between structures of different stoichiometries, the second-generation MLFF provides fairly accurate surface energies for all stoichiometries considered.
We then performed the PT runs using the updated MLFFs and found that the lowest energy structures of Py-CuD, Py-O$_\mathrm{v_{ss}}$ and CuD remain unchanged.\ This indicates that, in the present case, the MLFF trained only on a minimal stoichiometric configuration is sufficiently robust to enumerate low energy structures across {\em different but fixed} compositions of Cu$_{2}$O(111).\ Although an iterative refinement is certainly preferable (Supplementary Material S5 \cite{SupplementalMaterial}), the present simplistic approach is likely sufficiently robust for many cases.\

\subsection{\label{D} Summary of the structures }

Although the various structures may initially appear complex and confusing, our search of the phase space gives us a concise picture, which we summarize below.\

It is clear that the Cu$_\mathrm{cus}$ atoms are very unstable and tend to leave their ``original" position, creating a cavity between the Cu$_6$O$_6$ rings.\ However, this results in a very non-stoichiometric O-rich surface (CuD).\ Surfaces with higher Cu concentrations are obtained by agglomerating the Cu atoms within one of these cavities to form nanopyramids.\ These are always decorated with an O atom at the tip and connected to the bulk oxide  in a way that maximizes the number of Cu atoms with the preferred linear O--Cu--O coordination.\

When attempting to remove oxygen from the ST, nanopyramidal, or CuD structures, oxygen is preferentially removed from the subsurface layers.\ This behavior is consistently observed in the Cu$_\mathrm{4}$-ST, ST-O$_\mathrm{v_{ss}}$, CuD-O$_\mathrm{v_{ss}}$, Py-O$_\mathrm{v_{ss}}$ structures.\ The removal of subsurface oxygen leads to the formation of small Cu aggregates, which exhibit considerable stability, at least when evaluated using the PBE and r$^{2}$SCAN functionals.\

Similarly, Cu$_{6}$ aggregates can also form on the surface on top of three Cu atoms participating in the Cu$_6$O$_6$ rings [see Figure \ref{figure2}(b) and Figure \ref{figure4}(c)].\ This configuration is more Cu rich than the ``usual" Cu nanopyramids: The latter have one Cu atom without full O--Cu--O coordination beneath the apex, while the Cu$_{6}$ aggregates have three.

So we are left with remarkably few structural features: The ``usual'' and the Cu$_{6}$ nanopyramids, as well as subsurface oxygen vacancies, which similarly give rise to smaller Cu clusters.\ Other structural elements, such as undercoordinated Cu atoms remaining in the Cu$_6$O$_6$ rings cavities [Figure \ref{figure2}(d)], are a rare exception.\ Obviously, all these features change the stoichiometry in a rather complicated way, and only by constructing a surface phase diagram can we determine which of these structures will be sufficiently stable to appear in equilibrium conditions.\

\subsection{\label{E} Phase Diagram}

To identify the most favorable configuration among the various potential reconstructions obtained from PT simulations of ($\sqrt{3}$×$\sqrt{3}$)R30° and (2 $\times$ 2) supercells of the Cu\(_2\)O(111) surface, we calculated their relative surface energies ($\Delta$\(\gamma_{\text{slab}}\)) as a function of the oxygen chemical potential ($\mu_{\text{O}}$) via the following equation

\begin{equation} \label{eq:1}
\begin{split}
\Delta \gamma_{\text{slab}}( \mu_{\text{O}}) & = \frac{1}{2A} [\left(E^{\text{slab}} - E^{\text{ST}} \right) \\
       & - \Delta N_{\text{Cu}} (\frac{1}{2} E_{\text{Cu}_2\text{O}} -\frac{1}{4} E_{\text{O}_2} -\frac{1}{2}\mu_{\text{O}} )\\
      &  - \Delta N_{\text{O}} (\frac{1}{2} E_{\text{O}_2}  + \mu_{\text{O}})] 
\end{split}
\end{equation}

\noindent where \(E^{\text{slab}}\) is the energy of the slab for which the surface energy is being calculated, and \(E^{\text{ST}}\) represents the energy of the bulk-terminated stoichiometric surface configuration shown in Figure 2(a).\ The values \(\Delta N_{\text{Cu}}\) and \(\Delta N_{\text{O}}\) are the excess number of Cu and O atoms in the slab, respectively, and \(A\) denotes the surface unit cell area, with the factor \(\frac{1}{2}\) accounting for the presence of two identical surfaces of the slab.\ The chemical potential of O in the gas phase is noted as  \(\mu_{\text{O}}\) and is referenced to an isolated oxygen molecule O$_2$ (neglecting any entropy effects in the molecule).\ Its energy is
represented by \(E_{\text{O}_2}\).
The chemical potential of copper is assumed to be in thermodynamic equilibrium between
the gas phase oxygen and bulk $\text{Cu}_2\text{O}$, and hence given by 

\begin{equation} \label{eq:3}
\mu_{\text{Cu}} =  \frac{1}{2}[E_{\text{Cu}_2\text{O}} -(\frac{1}{2} E_{\text{O}_2} + \mu_{\text{O}})].
\end{equation}

It is emphasized that we have used the hard oxygen PAW potential (O$_{h}$) for calculating the energy of the O$_{2}$ molecule in Eq.\ (\ref{eq:1}), while employing soft oxygen PAW potentials (O$_{s}$) for all other calculations.\ This choice is driven by the superior ability of the hard potentials to retain an accurate description of short bonds such as in O$_2$ \cite{paier2005perdew}.\ The validity of this approach is supported by the following section \ref{sec:bulkphase}. 

\subsubsection{\label{sec:bulkphase} Bulk phase diagram}

The oxygen chemical potential at the phase boundary between bulk Cu$_{2}$O and CuO can be determined using the reaction:

\begin{equation} \label{eq:4}
 \text{Cu}_{2}\text{O} + \frac{1}{2} \text{O}_{2} \longrightarrow 2\text{CuO} . 
\end{equation}
The corresponding expression for the oxygen chemical potential ($\mu_{\text{O}}$) is given by:
\begin{equation} \label{eq:5}
  \mu_{\text{O}} = 2E_{\text{CuO}}-E_{\text{Cu}_2\text{O}}-\frac{1}{2}E_{\text{O}_{2}}.
\end{equation}

Adding and subtracting the oxygen atomic energy, one obtains 
\begin{equation} \label{eq:6}
\mu_{\text{O}} = 2E_{\text{CuO}}-E_{\text{Cu}_{2}\text{O}}-\frac{1}{2}E_{\text{O}_{2}} + E_{\text{O}}- E_{\text{O}}
\end{equation}
where $E_{\text{O}}$ denotes the energy of an isolated oxygen atom.
It is now possible to associate the two oxygen atomic energies to values calculated with different PAW potentials, and rearranging Eq.\ (\ref{eq:6}) further, one obtains:
\begin{equation} \label{eq:7}
\mu_{\text{O}} = (2E_{\mathrm{CuO}}^\mathrm{(s)}-E_{\mathrm{Cu}_{2}\text{O}}^\mathrm{(s)}- E_{\text{O}}^\mathrm{(s)})-(\frac{1}{2}E_{\text{O}_{2}}^{(h)}- E_{\text{O}}^{(h)})
\end{equation}

\noindent where the superscripts (s) and (h) stand for calculations using soft and hard oxygen potentials, respectively.\ As reference energies, we chose non-spin-polarized spherical atoms to calculate $E_{\mathrm{O}}^{(h)}$ and $E_{\text{O}}^{(s)}$.\ Note that not only the PAW potentials but also the energy cut-offs could be different for the calculations in the first and second parentheses of Eq.\ (\ref{eq:7}).\ In our calculations, we have used an energy cut-off of 1000 eV for computing $E_{\text{O}_{2}}^{(h)}$ and $E_{\text{O}}^{(h)}$.\ A detailed comparison of oxygen chemical potential values computed using all-soft, all-hard, and the mixed approach (as defined in Eq.\ \ref{eq:7}), is provided in Table \ref{Table2}.\\

\begin{table}
\caption{Oxygen chemical potentials (in eV) at the transition from Cu$_{2}$O $\rightarrow$ CuO and Cu $\rightarrow$ Cu$_{2}$O, for all-soft ($\text{O}_{s}$), all-hard ($\text{O}_{h}$), and mixed ($\text{O}_{h}$ for the O$_{2}$ molecule and $\text{O}_{s}$ for bulk CuO and Cu$_{2}$O) PAW potentials.}
\label{Table2}
\begin{tabular}{lcccccc}
\toprule
\multirow{3}{*}{\textbf{Method}} & \multicolumn{3}{c}{\textbf{Cu$_{2}$O} $\rightarrow$ \textbf{CuO}} & \multicolumn{3}{c}{\textbf{Cu $\rightarrow$ Cu$_{2}$O}} \\
\cmidrule(lr){2-4} \cmidrule(lr){5-7}
& \textbf{Soft} & \textbf{Hard} & \textbf{Mixed} & \textbf{Soft} & \textbf{Hard} & \textbf{Mixed} \\
& (eV) & (eV) & (eV) & (eV) & (eV) & (eV) \\
\midrule
PBE    & -1.474 & -1.081 & -1.085 & -1.524 & -1.160 & -1.135 \\
PBE+\textit{U}  & -1.451 & -1.067 & -1.062 & -1.777 & -1.406 & -1.388 \\
r$^{2}$SCAN & -1.941 & -1.495 & -1.487 & -1.865 & -1.428 & -1.411 \\
HSE06  & -1.670 & -1.175 & -1.168 & -1.946 & -1.438 & -1.444 \\
\midrule
Expt. \cite{chase1998nist}  & \multicolumn{3}{c}{-1.503} & \multicolumn{3}{c}{-1.751} \\
\bottomrule
\end{tabular}
\end{table}

It can be seen that the oxygen chemical potential values computed using the hard potentials ($\text{O}_{h}$) for both the O$_{2}$ molecule as well as bulk CuO and Cu$_{2}$O closely match those obtained with the mixed scheme ($\text{O}_{h}$ for the O$_{2}$ molecule and $\text{O}_{s}$ for bulk CuO and Cu$_{2}$O) across all functionals.\ In contrast, the soft potentials for all calculations lead to too negative binding energies for the oxides.\ This consistency confirms the mixed approach as a reliable and computationally efficient alternative to using hard PAW potentials throughout the calculations, which would dramatically increase the computational cost.\ It is noted that the isolated oxygen atom is best calculated in a spherical non-spin-polarized state, as errors in spin-polarized calculations are potentially more sizable for the soft PAW potential.\

For PBE and PBE+\textit{U}, the O$_{2}$ molecule is known to be overbound \cite{PhysRevB.73.195107} compared to oxides, so it is common to empirically reduce the binding energy, particularly in PBE+\textit{U}, where a shift of approximtely 400 meV would bring a good agreement with the experimental value in the present case.\ These corrections were estimated using the standard oxygen PAW potential provided with VASP \cite{PhysRevB.73.195107}.\ Highly accurate binding energies for small molecules close to Gaussian-type orbitals calculations can be only obtained using the hard PAW potentials \cite{paier2005perdew}.\ For instance, the standard oxygen potential underestimates the binding energy of O$_2$ by about 0.18 eV, the soft potential even by 0.77 eV.\ Accidentally, using O$_{s}$ leads to a fortuitous agreement with experiment.\ However, relying on such coincidence can obscure functional errors and makes comparison between different electronic structure codes impossible.\ 

Overall, PBE+\textit{U} yields reasonable results, particularly after applying the energy shift of 0.4 eV for the chemical potential.\ r$^{2}$SCAN partially corrects the overbinding of O$_{2}$ (or the underbinding of the oxides), but the improvement is limited, as Cu$_{2}$O is now unstable, in agreement with the materials project database \cite{Jain2013, ong2008li}.\ Meanwhile, HSE06 still exhibits molecular overbinding (or oxide underbinding), yielding results quite similar to PBE+\textit{U}.\ It is quite clear that even state-of-the-art functionals for the calculation of oxide formation energies remain imprecise, and some correction for chemical potentials is required.\ For PBE+\textit{U} and HSE, these are in the order of 300-400 meV.\

\subsubsection{\label{1} Surface phase diagram}
\begin{figure*}
\includegraphics[width= 15cm]{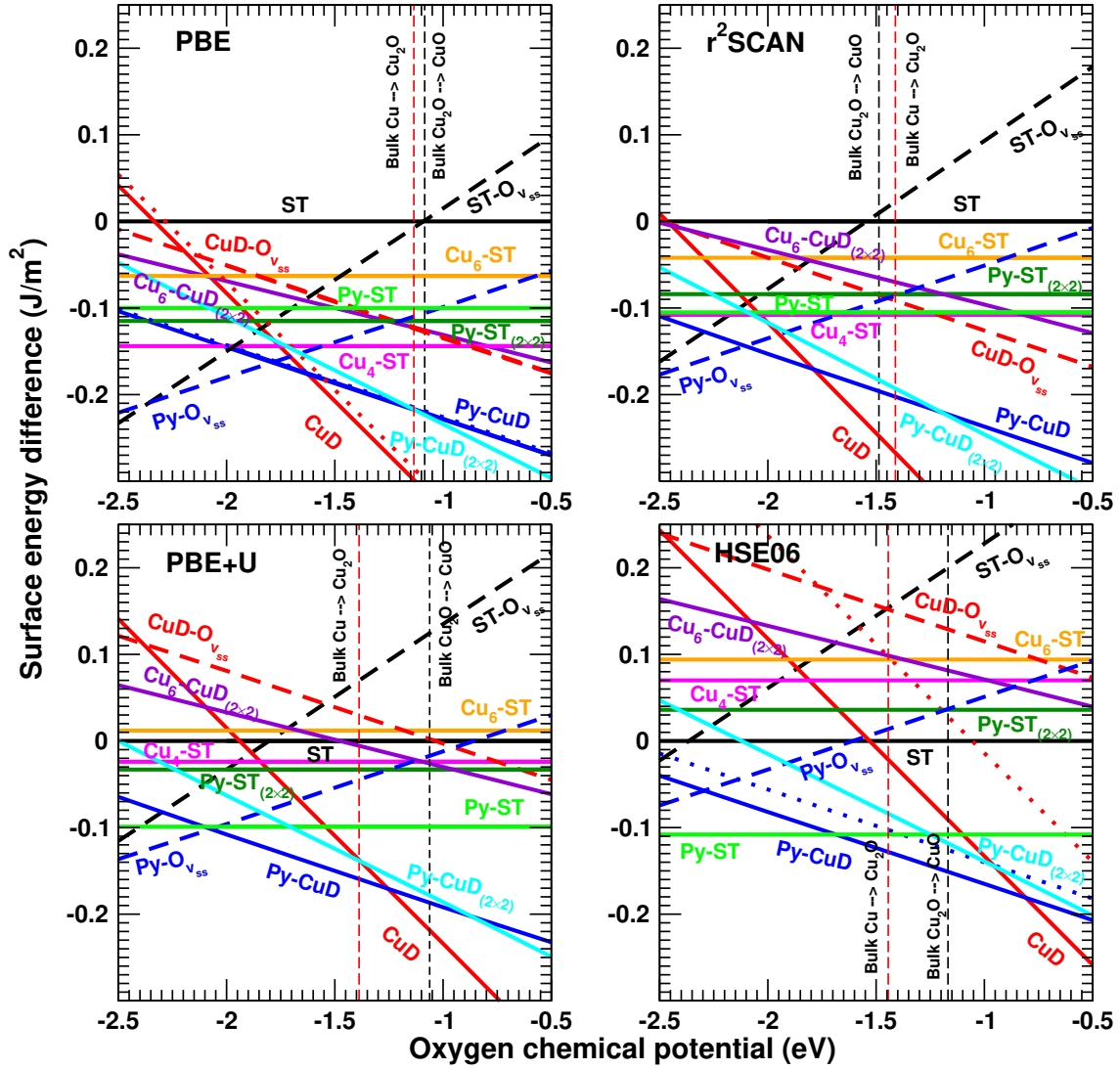}
\caption{\label{figure5}Surface energies of various possible reconstructions of Cu$_{2}$O(111) surface for ($\sqrt{3}$×$\sqrt{3}$)R30° and (2 $\times$ 2) supercells, as a function of oxygen chemical potential calculated using spin-polarized PBE, PBE+\textit{U}, r$^{2}$SCAN, and HSE06 functionals with mixed PAW potentials ($\text{O}_{h}$ for O$_{2}$ molecule and $\text{O}_{s}$ for bulk CuO and Cu$_{2}$O).\ Dashed red and blue lines in the PBE and HSE06 plots represent the surface energies of CuD and Py-CuD without spin-polarization.\ The most stable structures across different functionals, Py-ST, CuD, Py-CuD, and Py-O$_\mathrm{v_{ss}}$, correspond to Figure \ref{figure2}(d), \ref{figure3}(a), \ref{figure3}(d), and \ref{figure3}(e), respectively.\ Vertical black and red lines indicate the bulk transitions, from Cu$_{2}$O $\rightarrow$ CuO and Cu → Cu$_{2}$O, respectively.}
\end{figure*}

The surface energies of various possible reconstructions as a function of the oxygen chemical potential, calculated using Eq.\ (\ref{eq:1}), employing spin-polarized PBE, PBE+\textit{U}, r$^{2}$SCAN, and HSE06 functionals, are plotted in Figure \ref{figure5}.\ Additionally, surface energies calculated using MLFFs and FP (PBE) methods without spin-polarization are included in the Supplementary Material Figure S7 \cite{SupplementalMaterial}.\ 

We start with a comparison with previous work \cite{doi:10.1021/acs.jpcc.0c09330}.
In the present work, we have used the label Py-CuD for the nanopyramidal reconstruction to emphasize that this structure consists of nanopyramids on top of the CuD structure.\ In Ref. \onlinecite{doi:10.1021/acs.jpcc.0c09330}, this structure was labeled as Py.\ Using the hard oxygen potential for the oxygen molecule, the crossover points between the Py-CuD and CuD phases are $-1.55$~eV (PBE) and $-0.25~$eV (HSE06) for non-spin-polarized calculations (dashed red and blue lines in the PBE and HSE06 phase diagrams) and $-1.64$~eV (PBE) and $-0.80$~eV (HSE06) for spin-polarized calculations (solid red and blue lines).\ However, when using the soft oxygen potential for the oxygen molecule, these points shift to approximately $-1.96$~eV (PBE) and $-0.75$~eV (HSE06) for the non-spin-polarized case and $-2.0$~eV (PBE) and $-1.33$~eV (HSE06) for spin-polarized calculations, where the non-spin-polarized case aligns closely with the previous predictions \cite{doi:10.1021/acs.jpcc.0c09330}.\ This indicates that the use of the O$_{h}$ PAW potential for O$_{2}$ shifts the oxygen chemical potential to more positive values (by $\approx$ 0.4 eV), consistent with observations in the bulk phase diagram, see section \ref{sec:bulkphase}.\ We emphasize that other first-principles codes should give results consistent with the best PAW potentials; reported values using lower-quality PAW potentials are entirely code-specific.\ In addition, the inclusion of spin-polarization is crucial for non-stoichiometric cuprous oxide surfaces, or whenever copper atoms are in an oxidation state of +2, since the \textit{d}-shell becomes spin-polarized (\textit{d}$^9$ configuration for Cu).\

We now focus in more detail on the PBE results.\ Virtually, all the ML-predicted configurations are more stable than the ST configuration under PBE for a wide range of oxygen chemical potentials.\ This is not unexpected as the machine-learned potential was trained on PBE, and used to predict the stable structures.\ However, the convex hull is largely unchanged from previous studies \cite{doi:10.1021/acs.jpcc.0c09330,Gloystein_2021}.\ Specifically, the convex hull is formed by the CuD structure under oxidizing conditions, then a transition to the Py-CuD structure is observed, and finally, a subsurface oxygen vacancy forms under the nanopyramidal reconstruction.\ In our present study, we have extended the investigation to ($2\times 2$) supercells.\ The motivation for this is that in the experiment, the nanopyramids are not only observed in the ($\sqrt{3}\times\sqrt{3}$)R30° arrangement, but also seem to form less dense clusters \cite{doi:10.1021/acs.jpcc.0c09330}.\ However, all the structures we have found are somewhat higher in energy than the ($\sqrt{3}\times\sqrt{3}$)R30° Py-CuD structure, and nowhere do they lie below the convex hull. This suggests that, apart from kinetic limitations, the ($\sqrt{3}\times\sqrt{3}$)R30° Py-CuD structure should form easily on the surface.\ In the PBE (and r$^{2}$SCAN) calculations, Py-CuD is stable only at very reducing conditions, where the bulk would transform to metallic Cu in the thermodynamic equilibrium. 

Moving on to r$^{2}$SCAN, we see many similarities to PBE once we allow for a shift in the oxygen chemical potential of about $0.3-0.4~$eV. This is consistent with the bulk formation energy shifts.\ While individual structures also exhibit slight variations in surface energy, typically less than 0.02~J/m$^2$, the overall shape of the convex hull remains qualitatively unchanged.

In spite of the rough agreement between PBE and r$^{2}$SCAN, it must be doubted whether these functionals provide a satisfactory description of the surface energetics.\ For example, it is difficult to somehow translate the calculated chemical potentials to experimental values: one could try to adjust the chemical potentials so that the bulk formation energies agree with experiment, but PBE provides a stability regime for Cu$_2$O that is substantially too narrow (Table \ref{Table2}), and r$^{2}$SCAN even reverses the Cu$\rightarrow$Cu$_{2}$O and Cu$_{2}$O$\rightarrow$CuO transitions, which makes Cu$_2$O thermodynamically unstable.\ Clearly, the main problem is that without \textit{U} or using hybrid functionals one gets an unreasonable description of the energy difference between Cu$^{0}$, Cu$^{1+}$, and Cu$^{2+}$.\ Cu$^{2+}$ is characterized by a \textit{d}-shell occupied by 9 electrons, 5 in the majority and 4 in the minority spin-channel.\ Even state-of-the-art semilocal or meta-GGA functionals fail to capture this transition and the energy changes.\ 

PBE+\textit{U} should solve most of these problems.\ It is the only functional that predicts a crossing point from CuD to the nanopyramidal reconstruction within the stability range of Cu$_2$O.\ Specifically, PBE+\textit{U} predicts an intersection between Py-CuD and CuD at $-1.25~$eV, which after correction for the overbinding of the O$_2$ molecules shifts to about $-1.6~$eV, in good agreement with the experimental estimates of $\approx$ $-1.5$~eV. The other observations are as follows: all structures with subsurface oxygen vacancies become less stable except Py-O$_\mathrm{v_{ss}}$. This is consistent with \textit{U} destabilizing metallic Cu. Indeed, if we look at the Table \ref{Table2}, we see that \textit{U} only affects the transition from Cu to Cu$_2$O, leaving the transition from Cu$_2$O to CuO unchanged. In metallic Cu, the \textit{U} pushes the d-states down in energy and reduces the hybridization with \textit{s}-states and with neighboring \textit{d}-orbitals.\ Consistent with this, structures that contain metal clusters or oxygen sub-vacancies (leading to Cu-Cu bonds) are all destabilized by \textit{U}.\ This pushes the ST termination down in energy. For example, both Py-CuD and CuD are less stable (and much closer to the ST structure), but not enough to destabilize them. 
The last thing to note is that the Py-ST structure and Py-O$_\mathrm{v_{ss}}$ intersect the Py-CuD structure at about the same chemical potential, with the Py-O$_\mathrm{v_{ss}}$ dominating in the reduced state.\ As this is outside the stability regime of Cu$_2$O, it is probably of little practical relevance.\

We see that from PBE+\textit{U} to HSE06 these trends continue to some extent.\ While the transition point from Py-CuD to CuD was too negative for PBE and r$^{2}$SCAN (about $-1.6~$eV and $-1.8~$eV, respectively), it was at about the right energy for PBE+\textit{U} ($-1.2~$eV); with HSE06 it moved too much towards the positive side ($-0.8$~eV).\ An additional observation is that all structures containing oxygen vacancies become even more destabilized than with PBE+\textit{U}.\ It is somewhat difficult to reconcile these results within a simple picture, but overall the hybrid functional strongly favors the Py-CuD and Py-ST structures, disfavours sub-surface vacancies, as well as other structures involving metal clusters.\ Unfortunately, compared to the experiment, it stabilizes the nanopyramids too much, so that they dominate the phase diagram over the whole stability regime of Cu$_2$O.\ It is likely that one would need to reduce the amount of exact exchange to achieve better agreement with the experiment for the nanopyramidal stability regime.\ As a downside, a lower contribution of exact exchange will likely reduce the width of the stability regime of bulk Cu$_2$O, which happens to be accurately described with the HSE06 standard value of 25\% exact exchange used in the present work (Table \ref{Table2}).\
Overall, the description of the phase stability of complex multi-valent oxides remains very challenging.\ 

\section{Conclusion}
We have employed machine-learned force fields (MLFFs) using an on-the-fly methodology to explore various possible reconstructions of the Cu$_{2}$O(111) surface for both ($\sqrt{3}\times\sqrt{3}$)R30° and (2$\times$2) supercells.\ Using parallel tempering simulations, we successfully reproduced all previously reported configurations, including the nanopyramidal reconstruction, with a topographic height closely matching those observed in scanning tunneling microscopy (STM) images.\ In addition, we discovered a stoichiometric nanopyramidal configuration and two Cu cluster configurations within the ($\sqrt{3}$×$\sqrt{3}$)R30° unit cell, along with similar stoichiometric and Cu-deficient nanopyramidal arrangements and a Cu cluster configuration within the (2$\times$2) supercell.\ The occurrence of different variants of nanopyramidal reconstructions as the lowest-energy structures at many surface compositions may also explain the stability of CuO$_\mathrm{x}$-terminated tips in non-contact atomic force microscopy \cite{mönig2016submolecular}.\ We consider it likely that the structure of these tips is very similar to the pyramids of Cu$_{2}$O(111), with apical oxygen and linear O--Cu--O coordination of the Cu atoms below.\

We have also calculated the surface energies as a function of the oxygen chemical potential using spin-polarized PBE, PBE+\textit{U}, r$^{2}$SCAN and HSE06 functionals to determine the energetically most favourable reconstruction.\ Our analysis indicates that PBE+\textit{U} appears to be closest to experiment and is therefore, in the present case, the most ``reliable" functional for assessing the intersection between the Cu-deficient and nanopyramidal configurations.\ Clearly, predicting the stability of 3D transition metal oxides remains a major hurdle for density functional theory, one that cannot yet be fully addressed without empirically tunable parameters. This may come as a surprise, but it is certainly related to the fact that copper is a multivalent oxide, where the oxidation state Cu$^{2+}$ is associated with a hole polaron localized on a Cu atom. Most density functionals do not describe this energy difference and the associated energy gain or cost for localization well. In particular, we find that r$^{2}$SCAN does not provide a description in agreement with the experiment, but rather fails already at the level of the bulk oxides, with bulk Cu$_2$O being unstable at any oxygen partial pressure.\ Only hybrid functionals or the inclusion of a Hubbard U lead to a good agreement with experiment. However, as mentioned above, HSE seems to stabilize the nanopyramidal reconstruction too much, resulting in a surface diagram that is incompatible with the experiment.\

Overall, our results demonstrate that on-the-fly trained force fields, combined with parallel tempering, provide an efficient approach to generate complex surface reconstructions observed in experiments with minimal human intervention.\ In particular, our work aims to establish that even the simplest conceivable protocol yields a reasonable first surface diagram when the energies are evaluated a posteriori using density functional theory. Improvements to the training set are, however, also quite straightforward. Here, we added additional structures at two additional stoichiometries from the parallel tempering simulations.\ This improved second-generation machine-learned force field yields a surface diagram that closely matches the DFT-PBE surface phase diagram.\ In fact, it is obvious that the MLFF could be improved iteratively by adding structures generated by the simulations using the previous generation. Since the first-principles code that was applied uses a reliable Bayesian criterion to skip first-principles calculations, such a protocol will automatically skip redundant structures or duplicates, since their Bayesian error is small.\ In the present case, however, the second-generation MLFF was already good enough and only confirmed all structures generated by the first-generation force field.\ Furthermore, we expect our findings to remain valid across all types of machine-learning interatomic potentials, with predictions likely improving when employing message-passing networks.


\section*{acknowledgments}
The calculations were carried out on the Vienna Scientific Cluster (VSC). This work was supported by the Austrian Science Fund (FWF) through the SFB TACO project (Project No. 10.55776/F8100).

\section*{Data Availability}
The supporting data for this study is openly available at \cite{payal_dataset_2025}.

\nocite{*}

\bibliography{manuscript}

\end{document}


\preprint{APS/123-QED}

\title{Supplementary Material: Machine Learning Study of Surface Reconstructions of the Cu$_{2}$O(111) Surface}

\author{Payal Wadhwa}
\email{payal.wadhwa@univie.ac.at}
\email{payal\_wadhwa@iiitvadodara.ac.in}
\altaffiliation{\protect\\Present address: Indian Institute of Information Technology Vadodara, Gandhinagar-382028, Gujarat, India}
\affiliation{%
Faculty of Physics and Center for Computational Materials Science, University of Vienna, Kolingasse 14-16, A-1090 Vienna, Austria
}%

\author{Michael Schmid}
\affiliation{Institute of Applied Physics, TU Wien, A-1040 Vienna, Austria}
\author{Georg Kresse}%
 \affiliation{%
Faculty of Physics and Center for Computational Materials Science, University of Vienna, Kolingasse 14-16, A-1090 Vienna, Austria
}
 \affiliation{
VASP Software GmbH, Berggasse 21, A-1090 Vienna, Austria}
%




\date{\today}

\maketitle

\section{Principle Component Analysis of the training set}
To evaluate the structural diversity present in the training datasets for the first and second generation MLFFs, we performed Principal Component Analysis (PCA) using stoichiometry-aware structural descriptors (Cu atoms first, followed by O atoms, with each species padded to the dataset maxima per species).\ For the first-generation training set, the PC1–PC2 projection [Figure \ref{PC1}(a)] for a single stoichiometry shows that the configurations spread across a broad, structured region of principal component space, rather than forming one compact cluster. The distribution includes several dense groups linked by curved pathways, indicating that the dataset captures both distinct structural motifs and continuous transformations such as strain effects or thermal rearrangements.\ This suggests that the sampling is not limited to a narrow part of configuration space but instead covers multiple structural families and intermediate states.\

For the second-generation MLFF, the PC1–PC2 projection [Figure \ref{PC1}(b)] reveals three clearly separated clusters corresponding to the three stoichiometries in the dataset: stoichiometric (132 Cu, 66 O), Cu-deficient (134 Cu, 68 O), and Cu-rich (134 Cu, 66 O) atoms.\ The primary axis of variation (PC1) separates these compositions, while the spread within each cluster along PC2 reflects structural diversity at fixed composition.\ This demonstrates that the dataset not only covers multiple stoichiometries but also samples a range of structural distortions for each composition.\\

Overall, the PCA results show that both generations of MLFF training data cover diverse regions of configuration space, with clear compositional clusters and significant variation within each composition. This demonstrates broad dataset coverage, supporting the robustness of the MLFFs and confirming that they were trained on structurally rich configurations rather than a narrowly sampled set.

\begin{figure*}[h]
\centering
\includegraphics[width=14cm]{PCA_plot_new.eps}
\caption{\label{PC1}Principle component analysis (PCA) plots for the (a) first-generation and (b) second-generation machine-learned force fields.}
\end{figure*} 

\section{Additional Stoichiometric configurations}

\begin{figure}[h]
\includegraphics[width = 16cm]{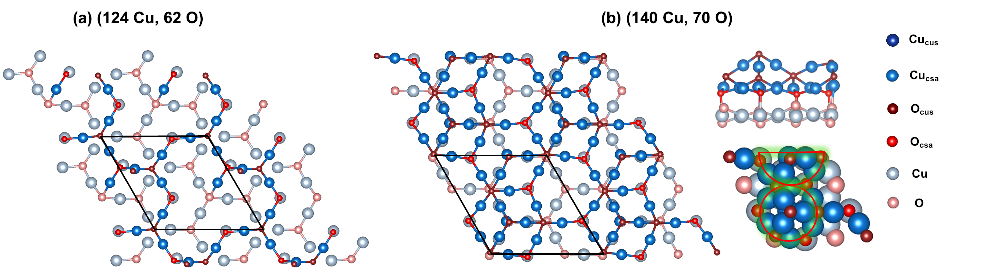}
\caption{\label{1}Atomic structures of the ($\sqrt{3}$×$\sqrt{3}$)R30° Cu$_{2}$O(111) surface showing stoichiometric configurations having (a) 124 Cu and 62 O atoms, (b) 140 Cu and 70 O atoms.\ All representations include the subsurface layer, with light blue and red spheres representing Cu and O atoms, respectively.\ In (b), the full and half red-green circles indicate the formation of complete and partial nanopyramids, respectively.\ The atom color coding is provided in the top right corner.}
\end{figure}

Surface reconstructions for stoichiometries containing (124 Cu, 62 O) and (140 Cu, 70 O) atoms are depicted in Figure \ref{1}.\ For the (124 Cu, 62 O) case, it is observed that the Cu$_6$O$_6$ hexagonal rings at the surface begin to break up.\ Meanwhile, for the configuration with (140 Cu, 70 O) atoms, one of the two Cu$_{6}$O$_{6}$ cavities, which are empty in the Py-CuD structure, becomes filled by 3 Cu atoms and a capping oxygen.\ Compared to a full-fledged nanopyramid, only one Cu atom is missing.\ This means any additional Cu atoms have a strong tendency to agglomerate in the Cu$_6$O$_6$ hexagonal rings. Since experimental STM images indicate the formation of only a single nanopyramid per ($\sqrt{3}$×$\sqrt{3}$)R30° superstructure cell \cite{doi:10.1021/acs.jpcc.2c04335,C8CP06023A}, no further stoichiometric cases are considered for the ($\sqrt{3} \times \sqrt{3}$)R30° Cu$_{2}$O(111) surface.

\section{Copper and Oxygen-deficient configurations}
\begin{figure}[h]
\includegraphics[width = 16cm]{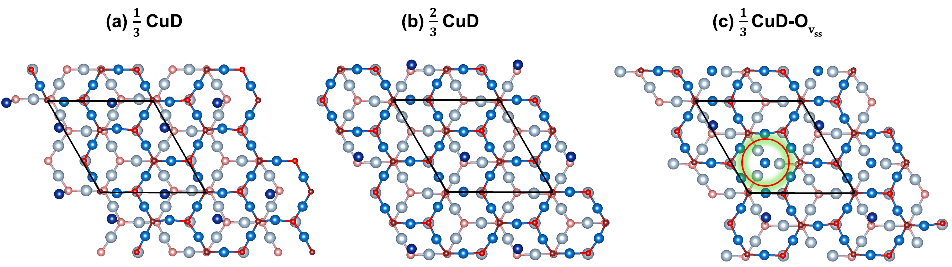}
\caption{\label{2}Atomic structures of the ($\sqrt{3}$×$\sqrt{3}$)R30° Cu$_{2}$O(111) surface showing (a) a one-third Cu-deficient surface ($\frac{1}{3}$CuD) and (b) a two-thirds Cu-deficient surface ($\frac{2}{3}$CuD), (c) a one-third Cu-deficient surface with an oxygen vacancy ($\frac{1}{3}$CuD-O$_\mathrm{v_{ss}}$).\ All representations include the subsurface layer, with light blue and red spheres representing Cu and O atoms, respectively.\ In (c), the subsurface oxygen vacancy is highlighted by a red-green circle.\ The atom color coding follows that of Figure \ref{1}.}
\end{figure}

Surface reconstructions for the $\frac{1}{3}$CuD, $\frac{2}{3}$CuD, and $\frac{1}{3}$CuD-O$_\mathrm{v_{ss}}$ configurations are illustrated in Figure \ref{2}. For the $\frac{1}{3}$CuD configuration, one-third of the Cu$_{6}$O$_{6}$ surface rings lack a Cu$_\mathrm{cus}$ atom, while for the $\frac{2}{3}$CuD configuration, the cavities of two-thirds of the Cu$_{6}$O$_{6}$ surface rings remain empty. Note that, unlike the stoichiometric surface (ST), which has three under-coordinated Cu$_\mathrm{cus}$ atoms, one or two Cu$_\mathrm{cus}$ atoms are insufficient to form a pyramid-like structure. In the experiment, the surface size is not limited to the ($\sqrt{3}$×$\sqrt{3}$)R30° superstructure cell; in this case, the collected Cu$_\mathrm{cus}$ from nearby cells would probably assemble into pyramids.\ 

In the $\frac{1}{3}$CuD-O$_\mathrm{v_{ss}}$ configuration, the oxygen vacancy migrates toward the subsurface layer, causing the Cu$_\mathrm{cus}$ atom to shift towards the center of the surface ring, forming a Cu-cluster similar to that is observed in the Cu$_\mathrm{4}$-ST configuration [circle in Figure \ref{2}(c)].\ Stability analysis reveals that the $\frac{1}{3}$CuD, $\frac{2}{3}$CuD, and $\frac{1}{3}$CuD-O$_\mathrm{v_{ss}}$ configurations are substantially less stable than CuD and Py-CuD configurations across all functionals, as depicted in Figure \ref{4}.\\

\begin{figure}[h]
\includegraphics[width = 16cm]{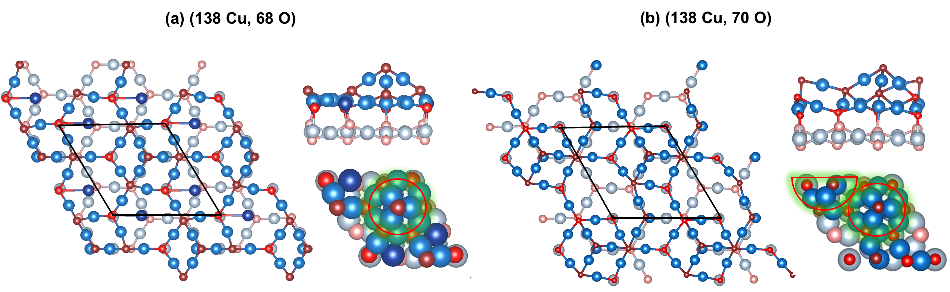}
\caption{\label{3}Atomic structures of the ($\sqrt{3}$×$\sqrt{3}$)R30° Cu$_{2}$O(111) surface showing configurations having (a) 138 Cu and 68 O atoms, (b) 138 Cu and 70 O atoms. All representations include the subsurface layer, with light blue and red spheres representing Cu and O atoms, respectively.\ The full and half red-green circles indicate the formation of complete and partial nanopyramids, respectively.\ The atom color coding follows that of Figure \ref{1}.}
\end{figure}

We also considered two additional stoichiometries, having (138 Cu, 68 O) and (138 Cu, 70 O) atoms.\ For the Cu-rich (138 Cu, 68 O) configuration, a nanopyramidal structure forms with two additional Cu$_\mathrm{cus}$ atoms as shown in Figure \ref{3}(a).\ In the (138 Cu, 70 O) configuration, the addition of one oxygen atom to the (138 Cu, 68 O)  configuration, drives the system closer to forming another nanopyramid, similar to (140 Cu, 70 O) configuration of Figure \ref{1}(b).\ The phase diagram indicates that all these configurations are unstable irrespective of the functional (PBE, PBE+U, r$^{2}$SCAN, or HSE06 functionals), as shown in Figure \ref{4}.

To determine the stability of the various reconstructions discussed above, their surface energies are calculated as a function of the oxygen chemical potential, employing spin-polarized PBE, PBE+U, r$^{2}$SCAN, and HSE06 functionals, as shown in Figure \ref{4}.\  To avoid overcrowding, the energies are only compared relative to the ST, CuD, CuD-O$_\mathrm{v_{ss}}$, Py-CuD, and Py-ST configurations.\

\begin{figure}[h]
\includegraphics[width=\textwidth]{phase_diagr_138_new.eps}
\caption{\label{4}Surface energies of various possible reconstructions of Cu$_{2}$O(111) surface for ($\sqrt{3}$×$\sqrt{3}$)R30° unit cell, as a function of oxygen chemical potential calculated using spin-polarized PBE, PBE+U, r$^{2}$SCAN, and HSE06 functionals.}
\end{figure}

\section{Energy errors for second-generation machine-learned force field}
For the second-generation machine-learned force field (MLFF) consisting of training sets of ST configuration as well as additional Py-CuD (Cu-deficient) and Py-O$_\mathrm{v_{ss}}$ (Cu-rich) configurations, the energy errors are shown in Figure \ref{5}(a).\ It is evident that the energies for different stoichiometries are harder to fit simultaneously, but the MLFF is accurate when restricted to a single stoichiometry.\ This suggests that the observed discrepancies could be dealt with by introducing an empirical energy shift corresponding to the stoichiometry as discussed in section \ref{S4}.\ Preferably, one can increase the weight of the energy equations during the fitting (Section \ref{S4}), although this slightly worsens the description of the forces, as discussed in the following.\\

\begin{figure}[h]
\includegraphics[width=14cm]{S5_new.eps}
\caption{\label{5}(a) Comparison of energies calculated using MLFF of the second-generation and first-principles (FP) calculations using PBE functionals without spin-polarization for ST, Py-{CuD}, and Py-O$_\mathrm{v_{ss}}$ configurations.\ Black points include the training set of the first-generation.\ All calculations for Cu-rich (Cu-deficient) structures have a Cu excess (deficiency) of one Cu atom per ($\sqrt{3}$×$\sqrt{3}$)R30° superstructure cell.\ (b) Root mean square error (RMSE) in energy for all configurations as a function of increasing energy weight.\ The inset shows an enlarged view of the plot.}
\end{figure}

\section{\label{S4}Phase diagram calculated using machine-learned force fields and first-principles}

The surface energies calculated using first and second generation MLFF and first-principles (FP) without spin-polarization are shown in Figure \ref{6}.\ As mentioned above, the surface energy curves of all the stoichiometries obtained from the first-generation MLFF are empirically shifted by a value that is solely based on the stoichiometry, represented by the slope of each curve.\ Specifically, an offset of \textit{s×b} is added to each curve, where \textit{b} is a constant common to all structures and \textit{s} is the slope of the curve.\ For shifting the surface energies obtained from the first-generation MLFF, the optimum value of \textit{b}  is determined by minimizing the mean squared error (MSE) between the corrected MLFF surface energies ($\gamma^\mathrm{MLFF^{corrected}}$) and the corresponding FP values ($\gamma^\mathrm{FP}$), given by:
\begin{equation}
 \mathrm{MSE} =  \frac{1}{n}\sum_{i=1}^{n}(\gamma_{i}^\mathrm{MLFF^{corrected}}-\gamma_{i}^\mathrm{FP})^2  
\end{equation}
One motivation for this empirical shift is as follows: As the oxygen content decreases, the number of Cu-Cu bonds increases, while the number of Cu-O bonds decreases. The MLFF needs to disentangle these contributions to the local energy. In the stoichiometric case, there is insufficient information to achieve this; however, when the stoichiometry is varied, the MLFF can correctly disentangle the energies, provided the energy weight is sufficiently large.

For the second-generation MLFF, we increased the weight of the energy equations during the fitting, to improve the accuracy of relative energy predictions across different stoichiometries.\ This adjustment was necessary because in the second-generation, the training set is dominated by stoichiometric (ST) configurations, with only a small number of added structures from Cu-deficient (Py-CuD) and Cu-rich (Py-O$_\mathrm{v_{ss}}$).\ Without reweighting, the model becomes biased toward the energy scale of the stoichiometric phase and fails to capture the relative energies of the non-stoichiometric configurations correctly.\ The value of the energy reweighting factor (set to 20 in our case) is determined by evaluating the root mean square error (RMSE) in energy across all stoichiometries, ensuring that the selected weight minimizes energy prediction errors across both stoichiometric and non-stoichiometric configurations [Figure \ref{5}(b)].\ As noted above, this slightly increases the force error, but with an increase of roughly 5\% the force errors remain acceptable.

\begin{figure}
\includegraphics[width=\textwidth]{Final_MLFF_FP_7nov.eps}
\caption{\label{6}Surface energies of various possible reconstructions of Cu$_{2}$O(111) surface for ($\sqrt{3}$×$\sqrt{3}$)R30° and (2 $\times$ 2) supercells, as a function of oxygen chemical potential calculated using first and second generation MLFF and FP within the PBE functional without spin-polarization.}
\end{figure}

From Figure \ref{6}, it can be seen that the first-generation MLFF overestimates the stability of ST-O$_\mathrm{v_{ss}}$ and Cu$_4$-ST compared to the pyramidal reconstructions and Py-O$_\mathrm{v_{ss}}$, however, the second-generation MLFF corrects this trend--- except for Cu$_4$-ST ---yielding a convex hull that is nearly consistent with FP calculations.\ For instance, the intersection point between Py-CuD and CuD using first and second generation MLLF are $-$1.82 eV and $-$1.67 eV, with the latter closely matching the FP value of $-$1.55 eV.\ The improved agreement of the second-generation MLFF with FP results is not surprising, as this model includes several non-stoichiometric configurations of Py-CuD and Py-O$_\mathrm{v_{ss}}$.\ However, these additions do not alter the lowest energy structures identified for specific stoichiometries using parallel tempering simulations with the first-generation MLLFs, as already discussed in the main manuscript. For instance, the ordering of the four stoichiometric structures is robust and agrees with the FP calculations even for the 1st generation MLFF.


\nocite{*}
